\documentstyle[prl,floats,aps]{revtex}

\begin{document}

\input{epsf.sty}

\draft

\twocolumn[\hsize\textwidth\columnwidth\hsize\csname 
@twocolumnfalse\endcsname

\title{Towards a Stable Numerical Evolution of Strongly Gravitating
Systems in General Relativity: The Conformal Treatments}

\author{
Miguel~Alcubierre${}^{(1)}$,
Bernd~Br{\"u}gmann${}^{(1)}$,
Thomas~Dramlitsch${}^{(1)}$,
Jos{\'e}~A.~Font${}^{(2)}$,
Philippos~Papadopoulos${}^{(3)}$,
Edward~Seidel${}^{(1,4)}$,
Nikolaos~Stergioulas${}^{(1,5)}$ and
Ryoji~Takahashi${}^{(1)}$ \medskip
}

\address{
$^{(1)}$Max-Planck-Institut f{\"u}r Gravitationsphysik,
Am M\"{u}hlenberg 1, D-14476 Golm, Germany \\
$^{(2)}$Max-Planck-Institut f{\"u}r Astrophysik,
Karl-Schwarzschild-Str. 1, D-85740 Garching, Germany \\
$^{(3)}$School of Computer Science and Maths, University of Portsmouth,
Portsmouth PO1 2EG, United Kingdom \\
$^{(4)}$ National Center for Supercomputing Applications,
Beckman Institute, 405 N. Mathews Ave., Urbana, IL 61801\\
$^{(5)}$ Department of Physics, Aristotle University of Thessaloniki,
Thessaloniki 54006, Greece 
}

\date{\today; AEI-2000-021}

\maketitle

\begin{abstract}
  
  We study the stability of three-dimensional numerical evolutions of
  the Einstein equations, comparing the standard ADM formulation to
  variations on a family of formulations that separate out the
  conformal and traceless parts of the system.  We develop an
  implementation of the conformal-traceless (CT) approach that has
  improved stability properties in evolving weak {\it and} strong
  gravitational fields, and for both vacuum {\it and} spacetimes with
  active coupling to matter sources.  Cases studied include weak and
  strong gravitational wave packets, black holes, boson stars and neutron
  stars.  We show under what conditions
  the CT approach gives better results in 3D numerical evolutions
  compared to the ADM formulation.  In particular, we show that our
  implementation of the CT approach gives more long term stable
  evolutions than ADM in all the cases studied, but is less accurate
  in the short term for the range of resolutions used in our 3D
  simulations.

\end{abstract}

\pacs{04.25.Dm, 04.30.Db, 97.60.Lf, 95.30.Sf}

\vskip2pc]

\narrowtext


\section{Introduction}
\label{sec:Introduction}

Three dimensional (3D) numerical relativity is an important technique
for exploring the strong field dynamics in realistic astrophysical
phenomena involving black holes and neutron stars.  It is expected to
play a role in analyzing gravitational waveforms to be observed soon,
one expects, with the new generation of gravitational wave detectors
going online worldwide in the next few
years~\cite{Flanagan97a,Flanagan97b}.  However, progress in 3D
numerical relativity, which has traditionally been based on the
Arnowitt-Deser-Misner (ADM)~\cite{Arnowitt62} system of evolution
equations, has been slow.  This is not only because of the immense
computational difficulties that 3D simulations represent, but to a
large extent it is due to severe instabilities often encountered
during such simulations.  Presently there is no complete understanding
of the causes of these instabilities in numerical evolutions of the ADM
equations.  This has prompted much recent effort in developing
alternative formulations of the 3+1 Einstein equations.

In this and a companion paper~\cite{Alcubierre99e} we focus on an
alternative approach based on a conformal decomposition of the metric
and the trace-free components of the extrinsic curvature.  The
conformal-tracefree (CT) approach was first devised by Nakamura in the
1980's in 3D calculations~\cite{Nakamura87,Nakamura89}, and then
modified and applied to work on gravitational waves~\cite{Shibata95},
and on neutron stars~\cite{Nakamura99a,Shibata99a}.  This approach was
not taken up by others in the community until a recent paper by
Baumgarte and Shapiro~\cite{Baumgarte99}, where a similar formulation
was compared with the standard ADM approach and shown to be superior,
in terms of both accuracy and stability, on tests involving weak
gravitational waves, with geodesic and harmonic slicing.  In a
followup paper, Baumgarte, Hughes, and Shapiro~\cite{Baumgarte99b}
applied the same formulation to systems with given (analytically
prescribed) matter sources, and found similar stability properties.
More recently fully hydrodynamical simulations employing the CT
approach have been reported in~\cite{Shibata99c,Shibata99d,Shibata99e}
in the context of collapse of rapidly-rotating (isolated) neutron
stars and coalescence and merger of binary neutron stars.  As we were
preparing this manuscript we have also become aware of work by Lehner,
Huq and Garrison~\cite{Lehner00a} where a comparison between the ADM
and CT formulation in spherical symmetry has been carried out in the
context of black hole excision.

In the companion paper~\cite{Alcubierre99e} we perform an analytic
investigation of the stability properties of the ADM and the CT
evolution equations.  Using a linearized plane wave analysis, we
identify features of the equations that we believe are responsible
for the difference in their stability properties.

In this paper we report the results of simulations of weak and strong
gravitational wave packets, black holes, boson stars and neutron stars
in various slicing conditions, including maximal slicing and a family
of algebraic slicings, and compare the results obtained by the ADM and
CT equations in different implementations.  We begin with a brief
presentation of the relevant equations in Sec.~\ref{sec:formulation}.
We then discuss the results of our numerical simulations in
section~\ref{sec:applications}. We consider vacuum spacetimes in
section~\ref{sec:vacuum}, and matter spacetimes in
section~\ref{sec:matter}.  In section~\ref{sec:waves}, we describe the
various implementations of the CT equations using gravitational wave
spacetimes as an example.  We identify two particular implementations,
which we call AFA and AF2, that give the best performance in long term
evolutions.  The essence of these implementations, is to ``actively
force'' (AF, see below) the trace of the conformally rescaled
extrinsic curvature (AFA), and for maximal slicing also the trace of
the extrinsic curvature (AF2), to zero in each step of the numerical
evolution.  In the sections that follow, we focus on comparing the AFA
and AF2 implementations to the results of the ADM equations for
evolutions of strong field systems including black holes, boson stars
and neutron stars.  We demonstrate that for this wide range of
systems, these two implementations of the CT equations always lead to
more stable long term evolutions.  However, we also find that for a
given resolution, the ADM results are often more accurate than the CT
results at early times, before the instabilities become apparent.  We
conclude with section~\ref{sec:conclusions}.  A study of the stability
properties of the iterative Crank-Nicholson (ICN) scheme, used for the
spacetime evolution of the simulations presented in this paper, can be
found in the Appendix.


\section{Formulation}
\label{sec:formulation}

We start reviewing briefly the formulations used for the
comparisons.

The standard ADM equations are\cite{York79}:
\begin{eqnarray}
\frac{d}{dt} \; \gamma_{ij} &=& -2\alpha K_{ij}, \label{metric evolution} \\
\frac{d}{dt} \; K_{ij} &=& -D_{i}D_{j}\alpha + \alpha \left(
\rule{0mm}{4mm} R_{ij} + K K_{ij} \right. \nonumber \\
& & \left.  - 2 K_{ik}K^{k}{}_{j} - {}^{(4)}R_{ij} \right),
\label{excurv evolution}
\end{eqnarray}
\noindent with
\begin{equation}
\frac{d}{dt} = \partial_t - \cal{L}_{\beta}
\end{equation}
and where $\cal{L}_{\beta}$ is the Lie derivative with respect to the
shift vector $\beta^i$. Here $R_{ij}$ is the Ricci tensor and $D_{i}$
the covariant derivative associated with the three-dimensional metric
$\gamma_{ij}$.  The 4-dimensional Ricci tensor ${}^{(4)}R_{ij}$ is
usually written in terms of the energy density $\rho$ and stress
tensor $S_{ij}$ of the matter as seen by the normal (Eulerian) observers:
\begin{equation}
{}^{(4)}R_{ij} = 8 \pi \left[ S_{ij} - \frac{1}{2} \left( S - \rho
\right) \right] .
\end{equation}

The conformal, trace-free reformulations of these equations make use of
a conformal decomposition of the three-metric, and the trace-free part
of the extrinsic curvature.  Here we follow closely the presentation
of Ref.~\cite{Baumgarte99}. The conformal three-metric $\tilde
\gamma_{ij}$ is written as
\begin{equation}
\tilde \gamma_{ij} = e^{- 4 \phi} \gamma_{ij},
\end{equation}
with the conformal factor chosen to be
\begin{equation}
e^{4 \phi} = \gamma^{1/3} \equiv \det(\gamma_{ij})^{1/3}.
\end{equation}
In this way the determinant of $\tilde \gamma_{ij}$ is unity.
The trace-free part of the extrinsic curvature $K_{ij}$, defined by
\begin{equation}
A_{ij} = K_{ij} - \frac{1}{3} \gamma_{ij} K,
\end{equation}
where $K = \gamma^{ij} K_{ij}$ is the trace of the extrinsic
curvature, is also conformally decomposed:
\begin{equation}
\tilde A_{ij} = e^{- 4 \phi} A_{ij}.
\end{equation}

So far, these are just definitions of new variables, with no clear
motivation for their introduction. Evolution equations for these new
quantities are easy to find, and we summarize here the
Baumgarte-Shapiro\cite{Baumgarte99} discussion on these equations, but
with an emphasis on the possible numerical implications of various
choices one can make.

The evolution equations for the conformal three--metric $\tilde
\gamma_{ij}$, and its related conformal factor $\phi$ are trivially
written as
\begin{eqnarray}
\frac{d}{dt} \tilde{\gamma}_{ij} &=& - 2 \alpha \tilde{A}_{ij} ,
\label{eq:evolg}
\\
\frac{d}{dt} \phi &=& - \frac{1}{6} \alpha K .
\label{eq:evolphi}
\end{eqnarray}

The evolution equation for the trace of the extrinsic curvature $K$
can easily be found to be
\begin{equation}
\frac{d}{dt} K = - \gamma^{ij} D_i D_j \alpha  + \alpha \left[
\tilde{A}_{ij} \tilde{A}^{ij} + \frac{1}{3} K^2 + \frac{1}{2}
\left( \rho + S \right) \right] ,
\label{eq:evolK}
\end{equation}
where the Hamiltonian constraint was used to eliminate the Ricci scalar.

For the evolution equation of the trace-free extrinsic curvature
$\tilde{A}_{ij}$ there are many possibilities.  A trivial manipulation
of Eq.~(\ref{excurv evolution}) yields:
\begin{eqnarray}
\frac{d}{dt} \tilde{A}_{ij} &=& e^{-4 \phi} \left[
 - D_i D_j \alpha + \alpha \left( R_{ij} - S_{ij} \right) \right]^{TF}
\noindent \\
&& + \alpha \left( K \tilde{A}_{ij} - 2 \tilde{A}_{il} \tilde{A}_j^l
\right) ,
\label{eq:evolA}
\end{eqnarray}
but as shown previously~\cite{Shibata95,Baumgarte99} there are many
ways to write several of the terms, especially those involving the
Ricci tensor.  For example, one could eliminate the Ricci scalar $R$
again through the use of the Hamiltonian constraint.

With the conformal decomposition of the three--metric, the Ricci
tensor now has two pieces, which we write as
\begin{equation}
R_{ij} = \tilde{R}_{ij} + R^{\phi}_{ij} .
\end{equation}
The ``conformal-factor'' part $R^{\phi}_{ij}$ is given directly by
straightforward computation of derivatives of $\phi$:
\begin{eqnarray}
R^{\phi}_{ij} &=& - 2 \tilde{D}_i \tilde{D}_j \phi - 2 \tilde{\gamma}_{ij}
\tilde{D}^l \tilde{D}_l \phi \noindent \\
&& + 4 \tilde{D}_i \phi \; \tilde{D}_j \phi - 4 \tilde{\gamma}_{ij}
\tilde{D}^l \phi \; \tilde{D}_l \phi ,
\end{eqnarray}
while the ``conformal'' part $\tilde{R}_{ij}$ can be computed in the
standard way from the conformal three--metric $\tilde \gamma_{ij}$.
To simplify notation, it is convenient to define what
Ref.~\cite{Baumgarte99} calls the ``conformal connection functions'':
\begin{equation}
\tilde{\Gamma}^i := \tilde{\gamma}^{jk} \tilde{\Gamma}^i_{jk} =
 - \tilde{\gamma}^{ij}_{~~,j} ,
\end{equation}
where the last equality holds if the determinant of the conformal
three--metric $\tilde \gamma$ is actually unity (notice that this should
be true analytically, but may not be numerically).

Using the conformal connection function, the Ricci tensor can be
written:
\begin{eqnarray}
\tilde R_{ij} & = & - \frac{1}{2} \tilde \gamma^{lm}
        \tilde \gamma_{ij,lm}
        + \tilde \gamma_{k(i} \partial_{j)} \tilde \Gamma^k
        + \tilde \Gamma^k \tilde \Gamma_{(ij)k} \nonumber \\
& &  + \tilde \gamma^{lm} \left( 2 \tilde \Gamma^k_{l(i}
        \tilde \Gamma_{j)km} + \tilde \Gamma^k_{im} \tilde \Gamma_{klj}
        \right).
\end{eqnarray}

Here again, one has choices in how the terms involving the conformal
connection functions $\tilde \Gamma^{i}$ are computed.  A
straightforward computation based on the Christoffel symbols could be
used (and usually is in standard ADM formulations), but this approach
leads to derivatives of the three--metric in no particular elliptic
form.  One would like to see an elliptic form as the principal part of
this expression, as it brings the $\tilde \gamma_{ij}-\tilde A_{ij}$
system a step closer to being hyperbolic.  Thanks to the definition of
the $\tilde \Gamma^{i}$'s, an explicitly elliptic operator is singled
out.  However, if the terms involving the $\tilde\Gamma^{i}$ are
evaluated directly in terms of derivatives of the three--metric, this
elliptic operator serves no special purpose, as other second
derivatives appear through derivatives of the $\tilde \Gamma^i$ which
spoils the elliptic nature of the operator as a whole.  If, on the
other hand, the $\tilde \Gamma^i$ are promoted to independent
variables, for which evolution equations can be derived, then the
expression for the Ricci tensor retains its elliptic character.  The
price to pay is that one must now evolve three additional quantities
in the evolution system.  Whether this has any numerical advantage
will depend on details of the implementation, and will be discussed
below.

Following this argument of promoting the $\tilde \Gamma^i$ to
independent variables, it is straightforward to derive their evolution
equation:
\begin{eqnarray}
\frac{\partial}{\partial t} \tilde \Gamma^i
&=& - \frac{\partial}{\partial x^j} \Big( 2 \alpha \tilde A^{ij}
- 2 \tilde \gamma^{m(j} \beta^{i)}_{~,m} \noindent \nonumber \\
&& + \frac{2}{3} \tilde \gamma^{ij} \beta^l_{~,l}
   + \beta^l \tilde \gamma^{ij}_{~~,l} \Big) .
\label{eq:evolGamma}
\end{eqnarray}

However, again there is a choice one can make in writing this
evolution equation; as pointed out in Ref.~\cite{Baumgarte99} it turns
out that the above choice leads to an unstable system. A choice which
will be shown to be better can be obtained by eliminating the
divergence of $\tilde A^{ij}$ with the help of the momentum
constraint:
\begin{eqnarray}
\frac{\partial}{\partial t} \tilde \Gamma^i
&=& - 2 \tilde A^{ij} \alpha_{,j} + 2 \alpha \Big(
\tilde \Gamma^i_{jk} \tilde A^{kj} \nonumber \\
&& - \frac{2}{3} \tilde \gamma^{ij} K_{,j}
- \tilde \gamma^{ij} S_j + 6 \tilde A^{ij} \phi_{,j} \Big)
\nonumber \\
&& - \frac{\partial}{\partial x^j} \Big(
\beta^l \tilde \gamma^{ij}_{~~,l}
- 2 \tilde \gamma^{m(j} \beta^{i)}_{~,m}
+ \frac{2}{3} \tilde \gamma^{ij} \beta^l_{~,l} \Big) .
\label{eq:evolGamma2}
\end{eqnarray}

With this reformulation, in addition to the evolution equations for
the conformal three--metric $\tilde \gamma_{ij}$~(\ref{eq:evolg}) and
the conformal-traceless extrinsic curvature variables $\tilde
A_{ij}$~(\ref{eq:evolA}), there are evolution equations for the
conformal factor $\phi$~(\ref{eq:evolphi}), and the trace of the
extrinsic curvature $K$~(\ref{eq:evolK}).  If the $\tilde \Gamma^i$
are promoted to the status of fundamental variables, as in
Ref.~\cite{Baumgarte99}, they can be evolved with~(\ref{eq:evolGamma2}).
(Note that the mixed first and second order evolution system for
$\{\phi, K, \tilde \gamma_{ij}, \tilde A_{ij}, \tilde \Gamma^i\}$ is
not in any immediate sense hyperbolic~\cite{FriedrichPrivateComm}).
In the original formulation of Shibata and Nakamura~\cite{Shibata95},
the auxiliary variables $F_i = -\sum_j \tilde \gamma_{ij,j}$ are used
instead of the $\tilde \Gamma^i$, and the final system of equations is
somewhat more complicated.

Ref.~\cite{Arbona99} shows that the CT system can also be interpreted
as a ``conformal second-order'' version of the Bona-Mass{\'o} system
with \mbox{$2 V_i = - (\tilde \Gamma_i + 8 \partial_i \phi)$}.


\subsection{Gauge}
\label{sec:gauge}

Systems of the CT type have been investigated with various slicing
conditions in the past.  The paper of Baumgarte and Shapiro considered
geodesic and harmonic slicing, while earlier work by Shibata and
Nakamura, and the more recent paper by Baumgarte, Hughes, and
Shapiro~\cite{Baumgarte99b} have also considered maximal slicing.
Here we have studied maximal slicing and a number of algebraic
slicings, and used them with different implementations of the CT
equations, on numerical evolutions of many different spacetimes.

Maximal slicing has the property that $K=0$, leading to an elliptic
equation for the lapse
\begin{equation}
\nabla^2 \alpha = \alpha \left[ K_{ij} K^{ij} + 4 \pi
\left( \rho + S \right) \right] .
\label{eq:maximal}
\end{equation}

Notice that in maximal slicing the evolution equations for
$\phi$ and $K$ become simply
\begin{equation}
d \phi/dt = 0,  \quad dK/dt = 0.
\end{equation}

The algebraic slicings that we will consider here correspond to the
family originally introduced by Bona and Mass\'{o}~\cite{Bona94b}, 
building on earlier work of Bernstein~\cite{Bernstein93a}
\begin{equation}
{\cal L}_t \; \alpha = - f(\alpha) \alpha^2 K,
\label{eq:BMslicing}
\end{equation}
with $f(\alpha)>0$ but otherwise arbitrary.  This family contains many
well known slicing conditions.  For example, taking $f=1$ we recover the
``harmonic'' slicing condition, which after a trivial integration
becomes
\begin{equation}
\alpha = F(x^i) + \gamma^{1/2} ,
\end{equation}
with $F$ an arbitrary function of space.  The name ``harmonic'' slicing
comes from the fact that it corresponds to the choice of a harmonic time
coordinate
\begin{equation}
\Box t = 0 .
\end{equation}

Another useful slicing condition is obtained by taking
\mbox{$f=N/\alpha$}.  This corresponds to the generalized ``1+log''
slicing condition~\cite{Arbona99} which after integration becomes
\begin{equation}
\alpha = F(x^i) + \log \gamma^{N/2} .
\end{equation}
(There is in fact some inconsistency in terminology as to whether the
$N=1$ or the $N=2$ case corresponds to the standard ``1+log'' slicing;
different choices being made by different authors.)

These type of algebraic slicings have an advantage over 
maximal slicing in terms of computational efficiency: It is much 
faster to integrate an evolution equation for the lapse than to solve 
an elliptic equation.  On the other hand, such algebraic slicings are 
prone to the development of gauge 
pathologies~\cite{Alcubierre97a,Alcubierre97b}.  The possibility of 
the appearance of such pathologies when using algebraic slicings 
should always be kept in mind, as a gauge pathology can easily be 
confused with a numerical instability: one can lose a lot of sleep 
trying to cure an ``instability'' that is in fact a true solution of 
our system of differential equations.

To finish discussing our choice of gauge, we need to mention the fact
that all the simulations described here have been carried out with the
shift vector set to zero.


\subsection{Boundary conditions}
\label{sec:boundaries}

In standard 3+1 numerical simulations, the computational domain covers
only a finite region of space.  One must therefore apply some sort of
artificial boundary condition at the edges of the numerical grid.
Ideally, one would like to find a boundary condition that does not
introduce numerical instabilities and allows gravitational waves to
leave the grid cleanly, with no artificial reflections.  This is in
itself a very difficult problem, since in the first place, there is no
local boundary condition that allows waves coming from any arbitrary
direction to leave the grid with no reflections, and second, there
does not even exist a clear way to define what a wave is in general
relativity except at asymptotic infinity.  In practice, what one looks
for is a condition that remains stable and allows some ``wave-like''
solutions to leave the grid without introducing large reflections at
the boundaries.  The amount of artificial reflection that results
typically depends on the specific form of the boundary condition, and
on the direction of motion of the wave fronts as they hit the
boundary~\cite{Gustafsson79}.

Since in this paper we are interested in the question of the stability
of the interior evolution, we will not worry too much about the
boundary conditions, and we will limit ourselves to describing a few
conditions that we have found to work well in practice.  The
conditions we have used are the following:

\begin{itemize}
  
\item Static boundary condition: The evolved variables are simply not
  updated at the boundary, and remain with their initial values there.
  This condition is very bad at handling waves since it reflects
  everything back in, but it can be very useful when studying
  situations that are supposed to remain static (as are some of the
  systems studied below), and where all the dynamics comes from
  numerical truncation errors.
  
\item Zero-order extrapolation or ``flat'' boundary condition: After
  evolving the interior, the value of a given variable at the boundary
  is simply copied from its value one grid point in (along the normal
  direction to the boundary).  This condition allows for some dynamics
  at the boundaries, and is somewhat better at absorbing waves than
  the static boundaries, but it still introduces a considerable amount
  of reflections.
  
\item Sommerfeld or ``radiative'' boundary condition: In this case we
  assume that the dynamical variables behave like a constant plus an
  outgoing radial wave at the boundaries, that is:
  \begin{equation}
  f(x^i,t) = f_0 + u(r-t)/r ,
  \label{eq:radiative1}
  \end{equation}
  where $r=\sqrt{x^2+y^2+z^2}$ and where the constant $f_0$ is taken
  to be one for diagonal metric components and zero for everything
  else.  The radiative condition assumes that the boundaries are in
  the wave zone, where the speed of light is essentially one, and
  where the gravitational waves behave as spherical wavefronts.  This
  boundary condition has been used before by other
  authors~\cite{Shibata95,Baumgarte99}, and it has been found that in
  practice it is very good at absorbing waves.
  
  It is in fact easier to implement a differential form of the
  radiative boundary condition than to use~(\ref{eq:radiative1})
  directly.  Consider a boundary that corresponds to a coordinate
  plane $x_i$=constant. Condition~(\ref{eq:radiative1}) then implies:
  \begin{equation}
  \frac{x_i}{r} \; \partial_t f + \partial_i f + \frac{x_i}{r^2} \;
  (f - f_0) = 0 .
  \label{eq:radiative2}
  \end{equation}
  One can now use simple finite differences to implement this last
  condition.  In our code we have implemented
  condition~(\ref{eq:radiative2}) consistently to second order in both
  time and space.
  
\item Robin boundary condition: This is a different type of
  ``extrapolating'' boundary condition, where one assumes that for
  large $r$ a given field behaves as:
  \begin{equation}
  f(x^i) = f_0 + k/r ,
  \label{eq:robin1}
  \end{equation}
  with $k$ constant.  This condition is clearly related to the
  radiative condition described above, but it contains no information
  about the time evolution.  Just as we did with the radiative
  condition, we in fact implement the Robin condition in differential
  form:
  \begin{equation}
  \partial_i f + \frac{1}{r} \; \left( f - f_0 \right) = 0 .
  \label{eq:robin2}
  \end {equation}
  The Robin boundary condition is usually better suited for solving
  elliptic problems than for use on dynamical variables.

\end{itemize}

Most of the simulations discussed below have been performed using the
radiative boundary condition~(\ref{eq:radiative2}) for the dynamical
variables, and the Robin boundary condition~(\ref{eq:robin2}) both for
constructing the initial data and for solving the maximal slicing
condition.  Whenever a different boundary condition is used, we say so
explicitely.


\section{Applications}
\label{sec:applications}

In this section we will apply the previous system of conformal
trace-free equations, exploring different implementations, in a series
of numerical experiments with different spacetimes.  The various
implementations we consider are:

\begin{itemize}
\item Promoting the $\Gamma$'s to independent variables.
\item Use the momentum constraints on the evolution equation for the
$\Gamma$'s.
\item Enforcing ${\rm tr}\tilde{A}=0$.
\item For maximal slicing, enforcing ${\rm tr}K=0$.
\end{itemize}

We will study the effects of these different implementations using
strong gravitational waves spacetimes.

All the numerical simulations presented here are carried out with the
Cactus code for numerical relativity co-developed in our
NCSA/Potsdam/Wash U collaboration and elsewhere.


\subsection{Vacuum Spacetimes}
\label{sec:vacuum}

We begin our discussion of the numerical simulations with vacuum
spacetimes in this subsection, examining the evolution of both strong
gravitational wave and black hole spacetimes.  In particular, we use
the gravitational wave simulations to illustrate the effects of the
various implementations of the CT approach.


\subsubsection{Pure Gravitational Waves}
\label{sec:waves}

We first turn to pure gravitational wave spacetimes.  The low
amplitude linear case has been studied, with a full 3D code, and
published previously, (a) in both the standard ADM formulation and the
Bona-Mass{\'o} hyperbolic formulation by~\cite{Anninos94d}, where no
fundamental differences were seen in performance at that time, and (b)
by Shibata and Nakamura~\cite{Shibata95} and Baumgarte and
Shapiro~\cite{Baumgarte99} in the CT approach as described above.  The
Baumgarte and Shapiro~\cite{Baumgarte99} work particularly showed the
strength of the CT formulation in the linearized case.  Here we extend
the study of these systems to include highly dynamic, strong field
regimes.  The study here is limited to tests that show the strengths
and weaknesses of the different formulations.  A study of the physics
of collapsing waves in full 3D numerical relativity is presented
elsewhere ~\cite{Alcubierre99b}.

We consider here a three--metric of the form originally considered by
Brill~\cite{Brill59}:
\begin{equation}
ds^2 = \Psi^4 \left[ e^{2q} \left( d\rho^2 + dz^2 \right)
+ \rho^2 d\phi^2 \right] =\Psi^4 \hat{ds}^{2},
\label{eqn:brillmetric}
\end{equation}
where $q$ is a free function subject to certain regularity and
fall-off conditions.  Different forms of the function $q$ have been
considered by different
authors~\cite{Holz93,Eppley79,Shibata97b,Gentle99}, but most work 
so far has concentrated only in constructing and analyzing the
initial data.

As in Ref.~\cite{Alcubierre99b}, we use a generalized form for the
function $q$, giving it a full 3D dependence,
following~\cite{Camarda97a,Allen98a,Brandt97a,Brandt97c}:
\begin{equation}
q = a \; \rho^2 \; e^{-r^2} \frac{(1 + c \rho^{2} \rm{cos}^{2}(m
\theta))}{(1+\rho^{2})},
\end{equation}
where $a$ and $c$ are constants, $r^2 = \rho^2 + z^2$ and $m$ is an
integer.  In this paper we focus on the axisymmetric case, $c=0$, for
simplicity, although using a non-zero value of $c$ does not affect the
results we discuss below.  All the runs discussed here where performed
using an iterative Crank-Nicholson (ICN) scheme with 3 iterations (see
appendix), and radiative boundary conditions.

The first case presented is an initial configuration with amplitude
$a$=4, corresponding to a strong wave, but not quite strong enough to
collapse to a black hole.  In the evolution of this data set the wave
implodes through the origin, oscillates a few times, and finally
disperses back to infinity leaving flat space behind, but in a
non-trivial spatial coordinate system~\cite{Alcubierre99b}.  The
evolution of this spacetime is highly non-linear, and the final
configuration has metric components with a large dynamical range.

\begin{figure}
\epsfxsize=3.4in
\epsfysize=2.5in
\epsfbox{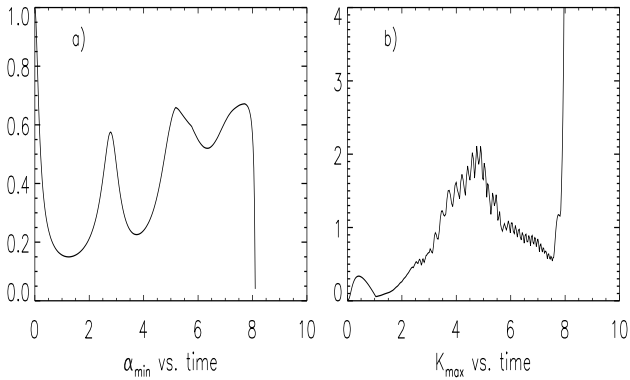}
\caption{
  a) Evolution of the minimum value of the lapse for an axisymmetric
  Brill wave data set with $a$=4, using the standard ADM formulation
  with maximal slicing. The simulation crashes at $t$=8 with a
  catastrophic collapse of the lapse. b) Evolution of the maximum
  value of the trace of the extrinsic curvature $K$.}
\label{fig:BW_ADM1}
\end{figure}

In Fig.~\ref{fig:BW_ADM1}a we show the evolution of the minimum value
of the lapse over the grid for a simulation done with the standard ADM
formulation, using maximal slicing, no shift and a radiative boundary
condition.  For this particular simulation we used a resolution of
$\Delta x$=0.08 and $67^3$ grid points.  Also, we used the fact that
our data is symmetric across coordinate planes to evolve only one
octant.  The simulation crashes at $t\simeq8$ when the lapse collapses
catastrophically in response to a blow up of the extrinsic curvature.
Fig.~\ref{fig:BW_ADM1}b shows the evolution of the maximum value of
the trace of the extrinsic curvature $K$.  Notice that even though we
are using maximal slicing, $K$ does not remain zero, and blows up
towards the end of the simulation.  The fact that $K$ does not remain
zero is not surprising, since the maximal slicing condition is solved
numerically, and thus a residual time derivative of $K$ is to be
expected.  The catastrophic blow-up, however, is a different matter
and points towards the existence of an unstable solution of our system
of equations.

\begin{figure}
\epsfxsize=3.4 in
\epsfysize=2.5 in
\epsfbox{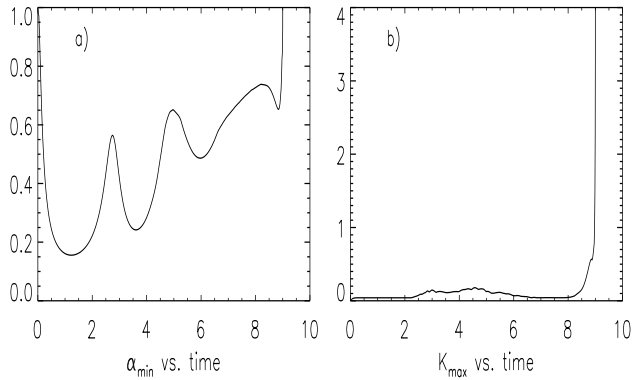}
\caption{
  a) Evolution of the minimum value of the lapse for an axisymmetric
  Brill wave data set with $a$=4, using the standard ADM formulation
  with maximal slicing and a K-driver. The simulation goes somewhat
  further, and now crashes at $t$=9 with a catastrophic blow-up of the
  lapse. b) Evolution of the maximum value of the trace of the
  extrinsic curvature $K$.  The trace now remains much smaller during
  the simulation.}
\label{fig:BW_ADM2}
\end{figure}

Fig.~\ref{fig:BW_ADM2} shows the same simulation, but now using the
so-called ``K-driving'' technique~\cite{Balakrishna96a}.  The idea
here is to add counter terms to the elliptic equation for the lapse to
drive the numerically produced non-zero $K$ (the trace of the
extrinsic curvature) back towards zero.  With K-driving, $K$ remains
much smaller until close to the point of crashing at $t$=9, with a
catastrophic blow-up of the lapse at the end.  This shows that a
better control of the time slicing is not enough to cure the
instability in the evolution: There exist unstable modes that are not
controlled by keeping the value of $K$ small.  (For an analysis of
possible unstable modes of the ADM equations, see~\cite{Alcubierre99e}.)

\begin{figure}
\epsfxsize=3.4 in
\epsfysize=5.5 in
\epsfbox{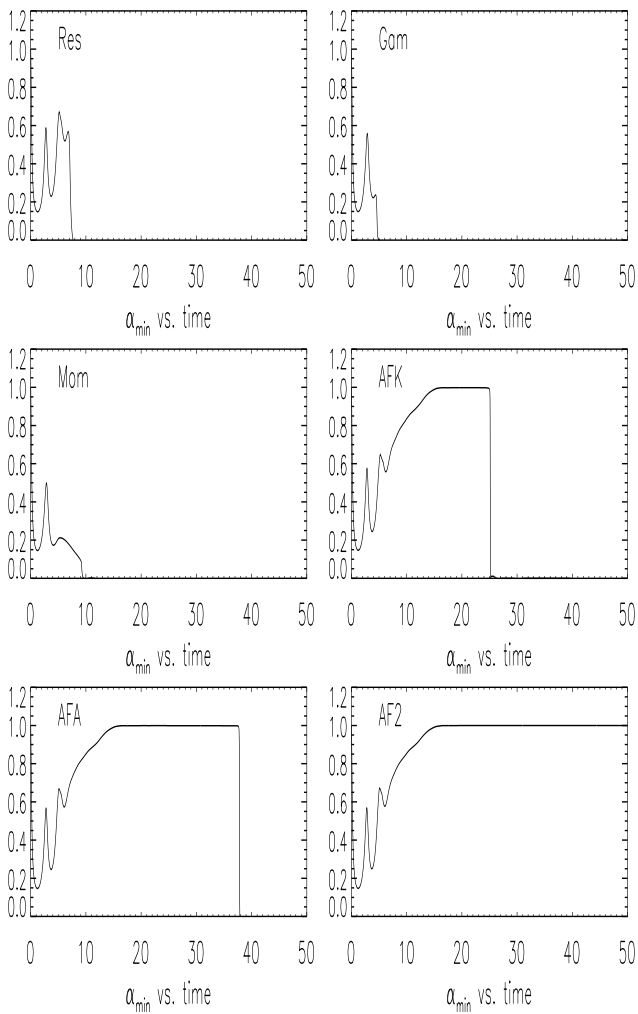}
\caption{Evolution of the minimum value of the lapse for an axisymmetric
  Brill wave data set with $a$=4, using the 6 different variations of
  the CT system described in the text.}
\label{fig:BW_ADMBS}
\end{figure}

Next, we show the evolution of the same system using again maximal
slicing, and different implementations of the CT formulation.  In
Fig.~\ref{fig:BW_ADMBS} we show again the central value of the lapse
for the same initial data.  The different runs correspond to the
following cases:

\vspace{5mm}

\begin{tabular}{ccccc}
& \, use of \, & \, use momentum \, & force & remove \\
& $\Gamma^i$   & constraints & \, $K$=0 \, & ${\rm tr} \tilde{A}$ \\
\\
Res & no  & -   & no  & no  \\
Gam & yes & no  & no  & no  \\
Mom & yes & yes & no  & no  \\
AFK & yes & yes & yes & no  \\
AFA & yes & yes & no  & yes  \\
AF2 & yes & yes & yes & yes \\
\end{tabular}

\vspace{5mm}

The first run uses the implementation denoted ``Res'' (for rescale).
It differs from the standard ADM equations only in the conformal
rescaling and the fact that $\phi$ and $K$ (which enter into the
evolution equation for $\tilde{A}_{ij}$) are now evolved separately.
The second run, with the implementation denoted ``Gam'' (for gamma),
introduces the $\Gamma^i$, but does not use the momentum constraints
to rewrite their evolution equations.  The third run uses the
implementation ``Mom'' (for momentum constraints) and represents a
straightforward coding of the the full set of CT
equations~\cite{Shibata95,Baumgarte99}, where the momentum constraints
are used to modify the evolution equations for the $\Gamma^i$, but
without adding anything else.  In the fourth run, which uses the
implementation ``AFK'' (for ``actively enforcing K''), we have forced
$K$ to remain zero by simply not evolving it, and we have also kept
$\phi$ time independent (see Eq.~(\ref{eq:evolphi})).  In the fifth
run we use the implementation ``AFA'', where we allow $K$ to evolve
freely, but actively force $\tilde{A}$ (the trace of $\tilde{A}_{ij}$)
to remain zero by subtracting it from $\tilde{A}_{ij}$ after each time
step:

\begin{equation}
\tilde{A}_{ij} \gets \tilde{A}_{ij} - \frac{1}{3} \tilde{\gamma}_{ij} \;
{\rm tr} \tilde{A} .
\end{equation}
And finally, in the sixth run we use the implementation ``AF2'' that
combines implementations AFK and AFA above by actively enforcing both
$K$=0 {\em and} $\tilde{A}$=0.  Notice that both $K$ and $\tilde{A}$
should be zero in principle in an exact evolution using the CT
equations with maximal slicing, but they do not remain so in actual
numerical evolutions unless actively enforced.
 
As can be seen from the figure, runs Res, Gam, Mom, AFK and AFA
eventually crash, but run AF2 with double active enforcement does not,
at least for the time scale under study.  The lapse returns to unity,
and the final static spacetime can be followed for a long time with no
sign of an instability (we have in fact followed run AF2 past $t$=100
and it still remains stable).  From the figure we also see that by
enforcing only $K$=0 or $A$=0 separately, as is done in runs AFK and
AFA, one still obtains improved stability, with the simulations
crashing at late times after the lapse has already returned to 1.
This shows that by enforcing only one of the two constraints, and
keeping the other options turned on, we still get a rather robust
system when compared to standard ADM.  Moreover, enforcing
$\tilde{A}$=0 appears to be more important than enforcing $K$=0, as
can be seen from the fact that run AFA crashes much later than run
AFK.

Finally, notice that run Gam crashes even sooner than run Res, which
shows that it is in fact better not to use the $\Gamma^i$ than to use
them without modifying their evolution equation.  For understanding
the need to use the momentum constraints in the CT approach, see the
companion paper~\cite{Alcubierre99e}.

\begin{figure}
\epsfxsize=3.4 in
\epsfysize=3.8 in
\epsfbox{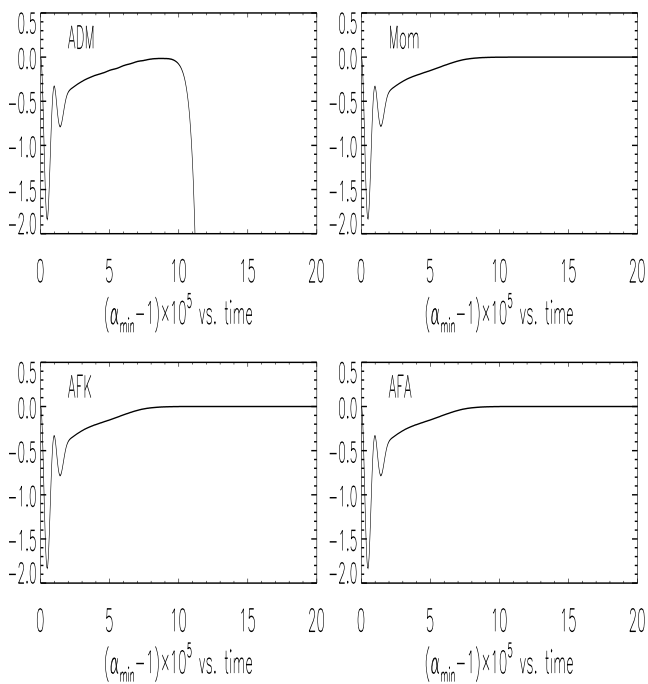}
\caption{Evolution of the minimum value of the lapse for an axisymmetric
  Brill wave data set with $a$=0.01 for the ADM system and three
  variations of the CT system (Mom, AFK, AFA).  Notice that since the
  lapse remains very close to 1, we are in fact plotting
  $(\alpha-1)\times10^5$.}
\label{fig:BW_weak}
\end{figure}

We note that the results found above for the different implementation
are generic for strong gravitational wave spacetimes, quite
independent of the precise parameter choices.  However, for weak
gravitational waves in the linear regime, the straightforward coding
of the CT equations (implementation ``Mom'') leads also to stable
evolutions as do the AFK, AFA and AF2 cases.  In
Fig.~\ref{fig:BW_weak} we show again the minimum value of the lapse
for the evolution of a wave with an amplitude of $a=0.01$, using the
ADM formulation and also the Mom, AFK and AFA implementations of the
CT system (since the lapse remains very close to 1, we are in fact
plotting $(\alpha-1)\times10^5$).  We see that while the ADM run
crashes at an early time ($t\simeq15$) with a catastrophic collapse of
the lapse, all three implementations of the CT equations give stable
evolutions and yield basically the same results for a weak wave.  We
have followed these three runs past $t$=100 with no instabilities
developing (the AF2 implementation is in fact just as stable, but we
don't include it in the figure).

From these studies (and many others with different parameters that we
have done) we can conclude that, for maximal slicing, the CT
formulation has better stability properties for the evolution of
strong field systems, as long as:

\begin{itemize}

\item The $\Gamma^i$ are promoted to independent variables.
  
\item The momentum constraints are used to transform the evolution
  equation for the $\Gamma^i$.  Evolving the $\Gamma^i$ without
  modifying their evolution equation is worse than not using them at
  all.
  
\item The trace of the extrinsic curvature $K$ is actively forced to
  be zero (the definition of maximal slicing).
  
\item The trace of the $\tilde{A}_{ij}$ is also actively forced to be
  zero.

\end{itemize}

\begin{figure}
\epsfxsize=3.4 in
\epsfysize=2.4 in
\epsfbox{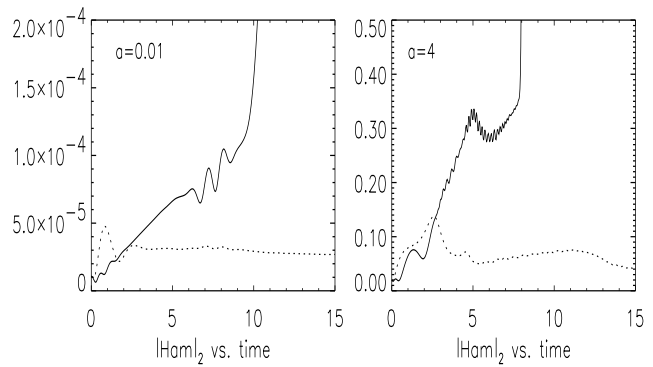}
\caption{
  L2-norms of the hamiltonian constraint for the $a$=0.01 and $a$=4
  cases, using the ADM (solid line) and the AF2 systems (dashed line).}
\label{fig:BW_ham}
\end{figure}

So far we have focused on the issue of long term stability.  Now we
want to compare accuracy of the CT and ADM formulations.  We
concentrate on the best implementation of the CT equations, the one we
labelled AF2.  In Fig.~\ref{fig:BW_ham} we show the L2-norms of the
hamiltonian constraint for the $a$=0.01 and $a$=4 cases discussed
above, using the ADM (solid line) and the AF2 systems (dashed line).
In both cases we see that for the ADM system, the L2-norm of the
hamiltonian constraint grows more or less linearly for some time (this
is more evident in the $a$=0.01 case) until just before the crash
when it blows up catastrophically.  In contrast, in the AF2 runs the
L2-norm of the hamiltonian constraint initially grows faster, but it
later settles on a constant value.  The fact that the ADM runs are
more accurate than the AF2 runs at early times appears to be quite
generic: we have found essentially the same behavior for all the
different parameters that we have studied.

\begin{figure}
\epsfxsize=3.4 in
\epsfysize=2.4 in
\epsfbox{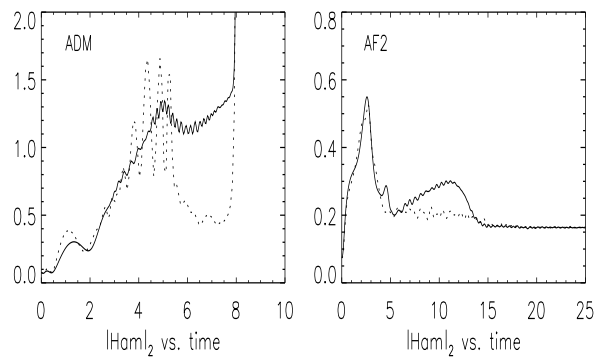}
\caption{
  Convergence of the L2-norms of the hamiltonian constraint for the
  $a$=4 case for both the ADM and the AF2 systems. The dashed lines
  show the L2 norm for a resolution for of 0.16, while the solid lines
  show the L2 norm for a resolution for of 0.08 multiplied by a factor
  of four.}
\label{fig:BW_conv}
\end{figure}

We have also performed convergence tests by running the same initial
data with different resolutions, and we have found that both the ADM
and AF2 evolutions are second order accurate.  As an example of this,
Fig.~\ref{fig:BW_conv} shows the L2-norms of the hamiltonian
constraint for both the ADM and the AF2 systems for two different
resolutions: The dashed lines show the L2 norm for a resolution of
$dx$=0.16 ($35^3$ grid points), while the solid lines show the L2 norm
for a resolution of $dx$=0.08 ($67^3$ grid points) multiplied by a
factor of four.  For second order convergence the solid and dashed
lines should fall on top of each other.  From the figure we see that
this is indeed true for most of the run in both cases.  For the ADM
run, second order convergence starts to fail shortly before the crash.
On the other hand, for the AF2 run we obtain slightly degraded
convergence (but still better than first order) for times between
$t$=5 and $t$=15 when the spacetime is very dynamic, indicating that
we haven't quite reached the second order convergence regime for the
resolutions considered here.

Though in this section we have concentrated in the case of maximal
slicing, we should mention that we have also performed many
simulations using the generalized ``1+log'' slicings.  The results are
in fact very similar to those reported here, except for the fact that
implementations AFK and AF2 can no longer be used (since $K$ in
non-zero for these slicing conditions).  We find that for these
algebraic slicings, implementation AFA is by far the best performer.

In the following subsections, we show that the above results on the
stability and accuracy of the ADM and CT systems are basically the
same for systems ranging from black holes to spacetimes coupled to
dynamical source fields.


\subsubsection{Black Holes}
\label{sec:blackholes}

Black holes have been the target of an intense research effort in
recent years in numerical relativity, and have proved particularly
difficult to handle in 3D evolutions. In the ``standard'' numerical
evolution of black holes using the ADM equations together with
singularity avoiding slicings, 3D simulations generally develop
instabilities and crash before $t=50M$, where $M$ is the mass of the
system~\cite{Anninos94c,Anninos96c,Bruegmann97}.  This falls far short
of the time required to model the complete inspiral of two black
holes, or even the head-on collision. Still, singularity avoiding
slicings combined with the ADM equations make it possible to evolve
through a brief part of the merger phase of two black holes with
momenta and spins, and from this point of view give the most generally
applicable method available. Future cures for grid stretching are
expected to be based on black hole
excision~\cite{Thornburg87,Seidel92a} or characteristic
slicings~\cite{Gomez98a}.

In the following we carry out a preliminary study of the CT
formulation in black hole evolutions with grid stretching.  It is
inevitable that the sharp peaks that develop in the metric function
due to grid stretching will cause the code to crash at some point in
the evolution. We consider the evolution of the Misner data as a
concrete example. The 3D numerical evolution of the Misner data in the
standard ADM setting with singularity avoiding slicing has previously
been studied using the so-called ``G''
code~\cite{Anninos96c,Camarda97a} and its
derivatives~\cite{Anninos94c}, developed by the NCSA/WashU group.
Comparable results for a single black hole can be found
in~\cite{Arbona99}.

\begin{figure}
\epsfxsize=3.2 in
\epsfysize=2.2 in
\epsfbox{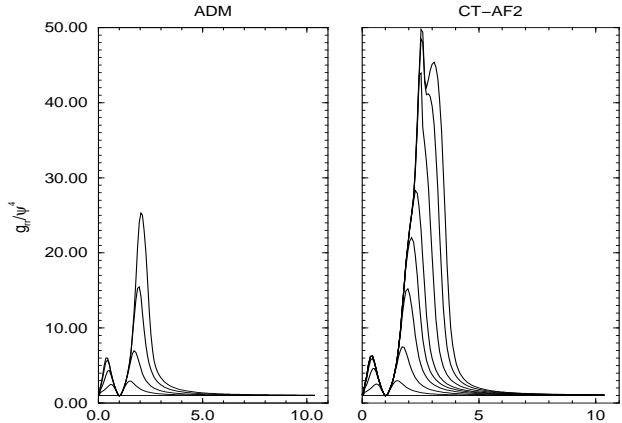}
\caption{Evolution of the radial-radial metric component along a line
  on the equatorial, plane at various times for Misner data
  ($\mu=2.2$).  Plots are every $3.5M$ in time.  The ADM system
  crashes after $t=14M$, while the AF2 system remains stable.}
\label{fig:Misner_grr}
\end{figure}

\begin{figure}
\epsfxsize=3.2 in
\epsfysize=2.2 in
\epsfbox{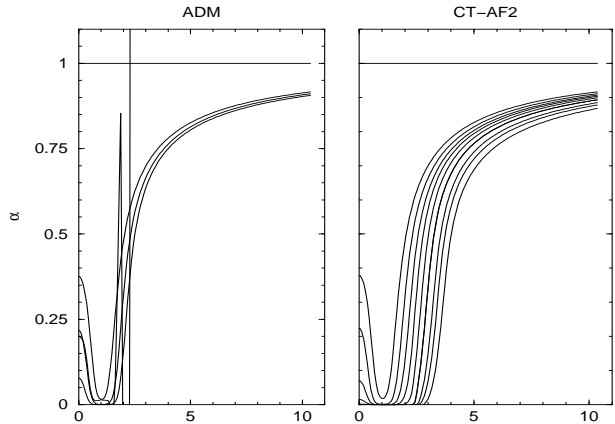} 
\vspace{3mm}
\caption{Evolution of the lapse function along a line on the
  equatorial plane at various times for Misner data ($\mu=2.2$).
  Plots are every $3.5M$ in time.  The ADM system crashes after $t=14M$,
  while the AF2 system remains stable.}
\label{fig:Misner_alp}
\end{figure}

In Fig.~\ref{fig:Misner_grr} and Fig.~\ref{fig:Misner_alp} we compare
the results of evolutions of Misner data with the separation parameter
$\mu=2.2$, corresponding to two initially well separated black holes,
on a grid of size $130^3$ with grid spacing $0.08$.  The only
difference in the simulations is the system of equations used to carry
out the evolution (ADM vs. AF2); all computational parameters, such as
parameters in the ICN finite differencing scheme, grid parameters,
radiative boundary conditions, and maximal slicing condition are the
same.  In Fig.~\ref{fig:Misner_grr}, first panel, we show the
radial-radial metric component along a line on the equatorial plane at
various times for the ADM case.  We can clearly see the familiar
ever-growing peak caused by the grid stretching associated with
singularity avoiding slicings. In the first panel of
Fig.~\ref{fig:Misner_alp} we show the lapse function along a line on
the equatorial plane at various times for the ADM case, and here an
instability becomes apparent at around $t=14M$ which is not yet
reflected in the metric.  This short wave length instability grows
rapidly and causes the code to crash at $t=14M$.  In the second panel
we show the AF2 case.  No metric instability is seen until towards the
end of the simulation at $t=24M$, although the peak appears to be
deformed.  At this time the radial metric function peak has grown to
about two times higher than that attained in the ADM case.  The lapse
for the AF2 case in Fig.~\ref{fig:Misner_alp} does not show an
instability.

\begin{figure}
\epsfxsize=3.2 in
\epsfysize=2.0 in
\epsfbox{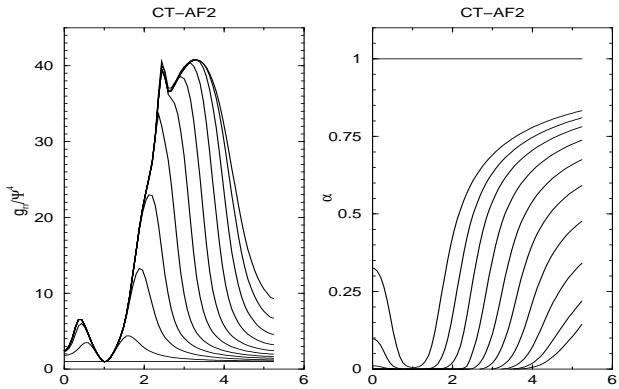} 
\vspace{3mm}
\caption{Evolution of the lapse and the metric at various times for
  Misner data ($\mu=2.2$). Plots are every $5M$ in time.
  With the AF2 system the evolution remains stable even
  after the metric peak is severely deformed.}
\label{fig:Misner_alp_grr}
\end{figure}

However, note that a smooth and stable evolution of the lapse does not
mean that the computed data is still useful. To emphasize this point,
Fig.~\ref{fig:Misner_alp_grr} shows the same run as above with AF2 on
a smaller grid with only $66^3$ points, but with the same grid spacing
as before (so the boundaries are much closer in). While ADM crashes when
the gradients in the metric become too severe, the AF2 run is able to
continue with a smooth lapse even after the metric becomes deformed
(cmp.~\cite{Arbona99} where the evolution of the metric is not
discussed).  The lapse eventually collapses in the whole grid,
freezing the evolution (so one could keep running ``forever'', but the
evolution becomes meaningless).

\begin{figure}
\epsfxsize=3.2 in
\epsfysize=2.5 in
\epsfbox{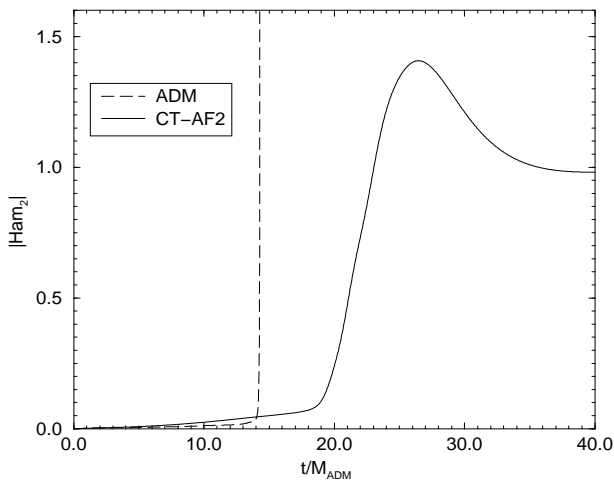} 
\caption{Evolution of the L2-norm of the Hamiltonian constraint for
  Misner data ($\mu=2.2$). The ADM system crashes at around $t=14M$,
  while the AF2 remains stable.  However, the accuracy of the AF2 run
  degrades significantly after around $t=20M$.}
\label{fig:Misner_ham_nm2}
\end{figure}

Next, we compare the accuracy of both simulations.  In
Fig.~\ref{fig:Misner_ham_nm2} we show the L2 norm of the Hamiltonian
constraint for a grid size of $130^3$.  The dashed line represents the
ADM run, and the solid line the AF2 run.  We see that the ADM results
are more accurate than the AF2 results until just after time $t=14M$,
when the instability in the ADM evolution begins to dominate and the
code crashes (with higher resolution this crash time can be delayed
somewhat). Starting at around $t=20M$ for AF2, there is a spurious
growth in the Hamiltonian constraint that corresponds to the
deformation in the metric. For maximal slicing one expects continuous
growth of a smooth metric peak, but with AF2 the shoulder in the lapse
seems to overtake the outward movement of the metric peak, freezing
its growth in an irregular manner.

These results for black holes with grid stretching cannot be compared
directly to the wave runs in the previous section because in the case
of the black hole runs we do not approach a static final state.
However, the CT formulation still offers some advantages over ADM in
achievable run time. We find stability far beyond were the runs are
meaningful, and it remains to be explored how far one can push the CT
runs while maintaining convergence.


\subsection{Matter Spacetimes}
\label{sec:matter}

In the previous sections we studied the stability properties of the
vacuum Einstein equations.  What will happen if these equations are
coupled to dynamical matter sources that are themselves governed by
evolution equations coupled to the spacetime geometry?  The complete
set of equations can now have more complicated types of unstable
modes.  What would be the effects of switching from the ADM
formulation to the CT formulation?

To respond to this question we consider next the following systems:
(i) the evolution of boson stars governed by the scalar field
Klein-Gordon equation and (ii) the evolution of neutron stars governed
by the hydrodynamical equations (general relativistic Euler
equations).  The numerical evolution of the Klein-Gordon equation is
straightforward with many well-known stable schemes. However, the
numerical evolution of the hydrodynamical equations is considerably
more challenging, especially in the presence of shocks or highly
relativistic flows.  For this purpose we use a recently developed
hydrodynamical code~\cite{font98b} which employs a conservative
formulation of the equations together with high-resolution
shock-capturing (HRSC) schemes based on approximate Riemann solvers.
In~\cite{font98b} we demonstrated that this code is capable of
handling hydrodynamical evolutions in a stable and accurate fashion
for a range of scenarios.

We focus here on analyzing the stability and accuracy of evolutions of
both static boson stars and static neutron stars using the ADM
formulation and the AFA implementation of the CT equations discussed
above.  We use the AFA implementation rather than AF2 because the
simulations discussed here have all been performed using algebraic
slicings and implementation AF2 applies only to maximal slicing.  The
main motivation for this has been the fact that, as we will show
below, implementation AFA with algebraic slicings already gives
excellent results when compared with standard ADM and is far less
computationally expensive than runs that use maximal slicing.


\subsubsection{Boson Stars}
\label{sec:bosonstars}

We begin with a simple kind of matter source: self-gravitating scalar
fields.  This system has served as a useful testbed for developing
numerical techniques for dealing with relativistic matter coupled to
the Einstein
equations~\cite{Seidel90b,Balakrishna96a,Balakrishna98a,Balakrishna98c},
and also has a distinguished history in the field, having provided the
first example of critical phenomena in relativity~\cite{Choptuik93}.

The dynamics of a massive scalar field are described by the minimally
coupled Klein-Gordon (KG) equation
\begin{equation}
\Box_g\phi=m^2\phi ,
\end{equation}
(see, e.g.~\cite{Seidel90b}). The KG equation can be obtained from the 
Lagrangian
\begin{equation}
{\cal L}= \frac{1}{2}g^{\mu\nu}\phi ,_{\mu}\phi^*,_{\nu} + 
\frac{1}{2}m|\phi|^2 ,
\end{equation}
which leads to the stress-energy tensor 
\begin{equation}
T_{\mu\nu}=\frac{2}{\sqrt{-g}}\frac{\delta {\mathcal L}}{\delta 
g^{\mu\nu}} ,
\end{equation}
which is used as the matter source for the Einstein equations.

Self-gravitating massive scalar fields have bound, star-like solutions
called boson stars with stability properties very much like those of
neutron stars.  These objects have been studied numerically,
extensively in 1D~\cite{Seidel90b,Balakrishna98c} and also in
3D~\cite{Balakrishna96a}. Apart from the fact that their evolution
equation is much simpler than the hydrodynamical equations, boson
stars are also easier to handle numerically when compared to neutron
stars because they have no sharp changes in the density distribution
near the surface layer of the star.  For more details on the
properties of boson stars and their behavior under perturbations
see~\cite{Seidel90b} and references cited therein.

We perform our numerical evolutions of boson stars by writing the
KG equation as a flux-conservative system of the form
\begin{equation}
\dot u_a = \partial_b F^b_a + S_a^bu_b
\end{equation}
where $\vec u$ contains the scalar field and its time and space
derivatives.  The method used to integrate this equation is a
symmetrized MacCormack with both directional and Strang splitting.
Symmetrized here means that the order of left-hand and right-hand
differencing changes every time step (this improves the stability of
the scalar field evolution).  The code for solving the KG equation
converges to second order in time and space.  See
Ref.~\cite{Dramlitsch99,Balakrishna99a} for details of the numerical
methods.

We have carried out evolutions of equilibrium boson star
configurations with the metric background held fixed artificially (not
updating the metric functions), and evolutions of the metric of such
configurations with the scalar field held fixed artificially (not
updating the scalar field), for a range of compactness of the boson
stars, using both the ADM and AFA schemes.  For all these cases, we
have seen that the simulations are stable and second order convergent.
The case of coupled spacetime-scalar field evolution is much more
challenging, and we focus on that below.

\begin{figure}
\vspace{-0.5cm}
\epsfxsize=3.3 in
\epsfysize=2.5 in
\epsfbox{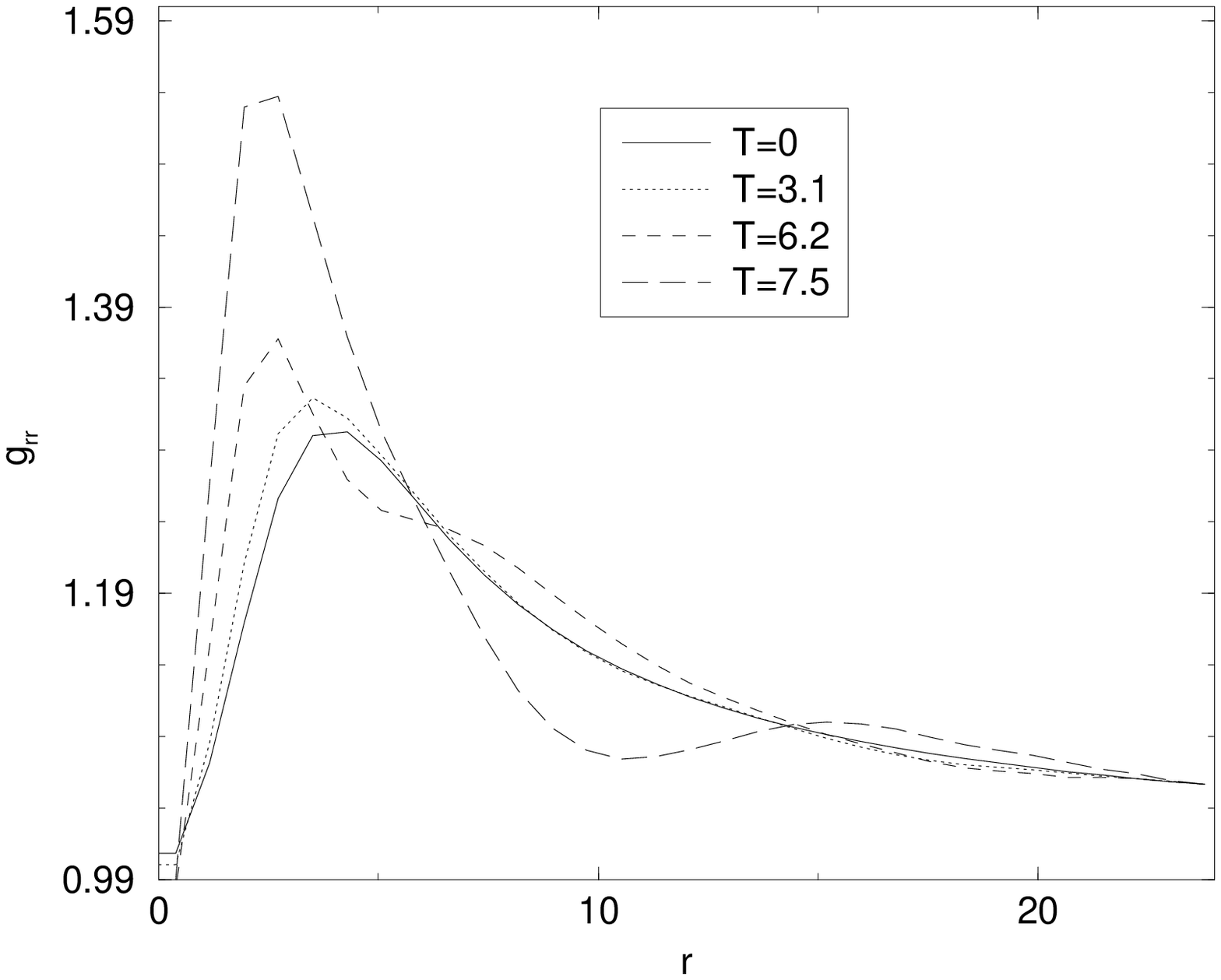}
\epsfxsize=3.3 in
\epsfysize=2.5 in
\epsfbox{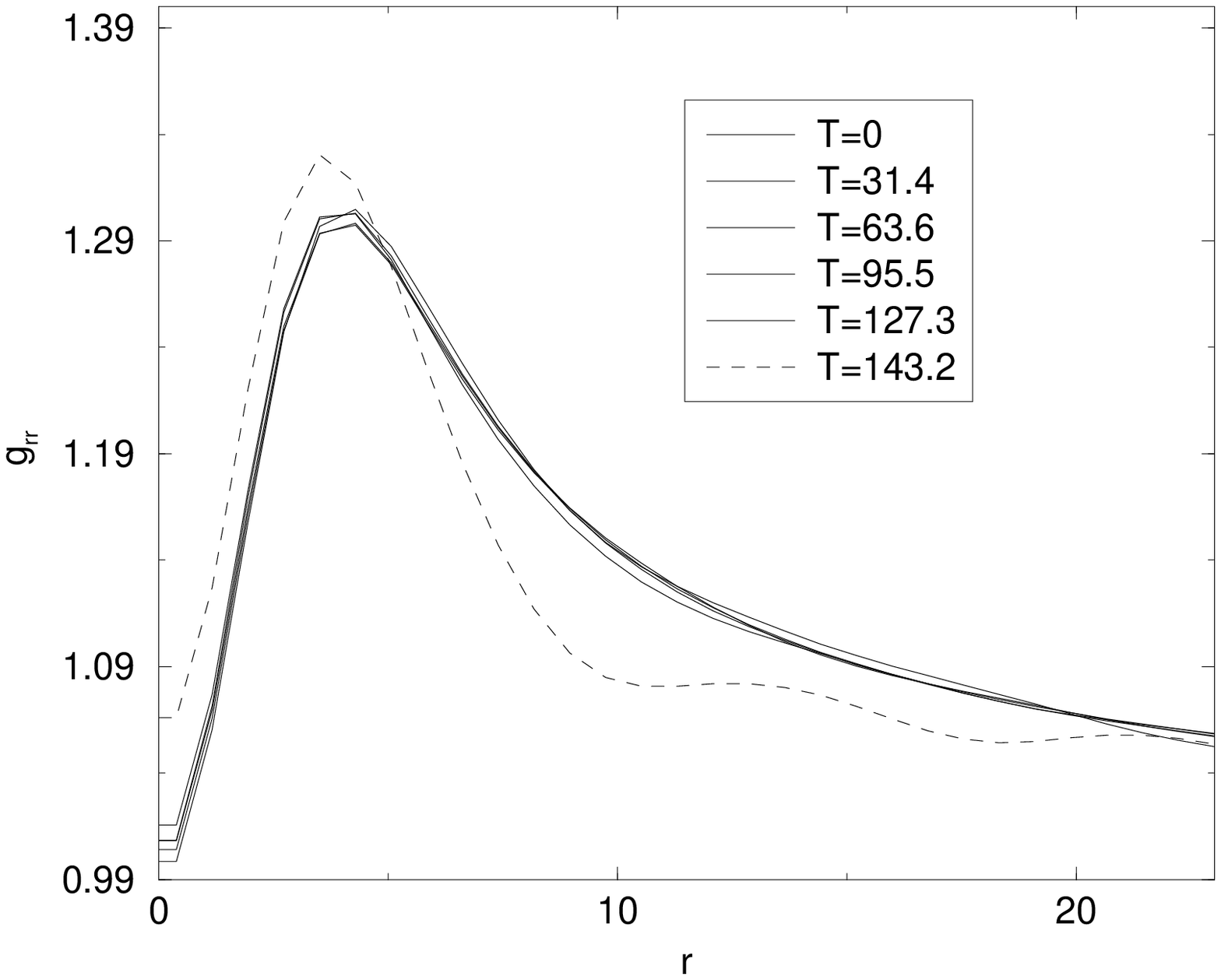}
\caption{Evolution of the radial metric function $g_{rr}$ using ADM
  (upper panel) and the AFA implementation (lower panel). The
  ADM evolution crashes at $t\simeq8$.}
\label{fig:scalar_eq_grr}
\end{figure}

We begin by showing an equilibrium boson star with a central density
near the maximum stable value (field strength at center $\phi_0=0.26$,
total mass \mbox{$M=0.6322~{m_p}^2/m$}, with $m_p$ the Planck mass,
$m$ the mass of the scalar field).  In Fig.~\ref{fig:scalar_eq_grr},
we show the evolution of radial metric component $g_{rr}$ in a fully
coupled simulation, using a three step ICN scheme, 1+log slicing with
$N=2$, no shift, a radiative boundary condition on the metric, and a
flat boundary condition on the scalar field. A $32^3$ grid is used to
cover only one octant.  In the first panel we show the results of the
ADM evolution.  We see that for a short time, the spacetime remains
nearly static (as it should).  However, a short wavelength instability
becomes significant by time $t$=7, and quickly grows causing the code
to crash.  The time $t$ here is expressed in terms of the intrinsic
oscillation time scale of the scalar field (the exact equilibrium
boson star field has the form $\psi(r)e^{i t}$).  In the second panel
we show the evolution with exactly the same setup but using now
implementation AFA instead of ADM.  We see that the static
configuration is maintained for a much longer time.  Towards the end
of the evolution, near $t=150$, we see that numerical error starts to
build-up near the boundary of the computational domain.

In Fig.~\ref{fig:boson_comp} below, we compare the L2-norm of the
hamiltonian constraint for the ADM and AFA runs.  We see that at early
times the ADM run gives a more accurate result, but instabilities
cause the L2-norm to blow by $t\simeq8$.  For the AFA run the
constraint violation is larger at first, but the evolution remains
stable or a much longer time.  The oscillation of the hamiltonian
constraint we see here can be understood as a reaction of the scalar
field to the numerical truncation error, which can be interpreted as a
kind of perturbation.  The frequency of these oscillations coincides
with the ones obtained in 1D studies of perturbed boson stars. Notice
that with the ADM run the code crashes so early that one can not even
see the first oscillation.

\begin{figure}
\vspace{-0.5cm}
\epsfxsize=3.3 in
\epsfysize=2.5 in
\epsfbox{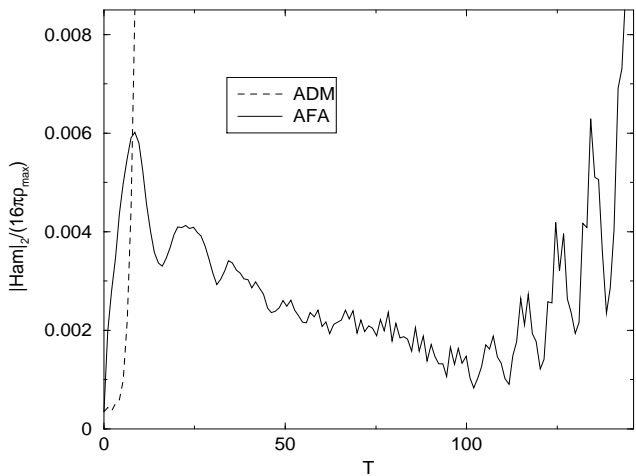}
\caption{Evolution of the L2-norm of the Hamiltonian constraint for
  a static and stable fully coupled boson star using the ADM and AFA
  systems.  The evolutions where carried out on a $32^3$ grid, with a
  resolution of $\Delta x=0.45$.  }
\label{fig:boson_comp}
\end{figure}


\subsubsection{Static Neutron Stars}
\label{sec:staticneutron}

We turn now to the study of hydrodynamical evolutions of neutron
stars. In~\cite{font98b} we developed a three-dimensional, fully
relativistic code to integrate the hydrodynamical equations coupled to
the ADM equations. Convergence studies using polytropic neutron stars
showed that the code is second order accurate in both space and time.
For the integration of the hydrodynamical equations we used HRSC schemes
of the total-variation-diminishing (TVD) class, with a
piecewise-linear reconstruction of a sufficient set of hydrodynamical
variables (rest-mass density, three-velocity and internal energy
density).  For more details on the schemes available in the code, see
\cite{font98b}.  In the studies reported in this paper we use the ICN
scheme for the integration of the spacetime equations (either ADM or
AFA) and Roe's approximate Riemann solver for the hydrodynamical
equations. We use ``1+log'' slicing with $N=2$.

As in the boson star studies we have first considered evolutions which
test separately the individual components of the code. In these, we
either solve the hydrodynamical equations in a prescribed (static)
spacetime or the gravitational field equations for a prescribed matter
source.  In particular, we have evolved static neutron star
configurations with a zero-temperature polytropic equation of state,
of the form $P=K\rho^{\Gamma}$ (where $P$ is pressure and $\rho$ is
rest-mass density).  This included stars with a large polytropic index
$\Gamma$ (very stiff) having density profiles with a discontinuous
first derivative at the surface. In the case of
prescribed matter sources, we have confirmed that the comparison of
the AFA and AF2 systems to the ADM system, in terms of
stability and accuracy, remains the same as in the vacuum cases
studied above.  Static neutron stars with polytropic index $\Gamma=2$
have also been studied in \cite{Baumgarte99b} using the CT equations
with prescribed hydrodynamical sources.

We focus next on the coupled spacetime and hydrodynamical evolution of
static Tolman-Oppenheimer-Volkoff (TOV)~\cite{Misner73} neutron stars
(in isotropic coordinates).  Again, we compare the results obtained
using the AFA implementation of the CT equations to those of the ADM
equations.  In principle both the matter distribution inside the star
and the spacetime should remain static.  In practice they evolve due
to truncation errors of the finite-difference scheme, with the
hydrodynamics and the spacetime responding to one another.  The static
TOV solution provides a reference to monitor the accuracy of the
coupled numerical evolution. Note that in these evolutions, static
outer boundary conditions were used.

\begin{figure}
\epsfxsize=3.3 in
\epsfysize=2.5 in
\epsfbox{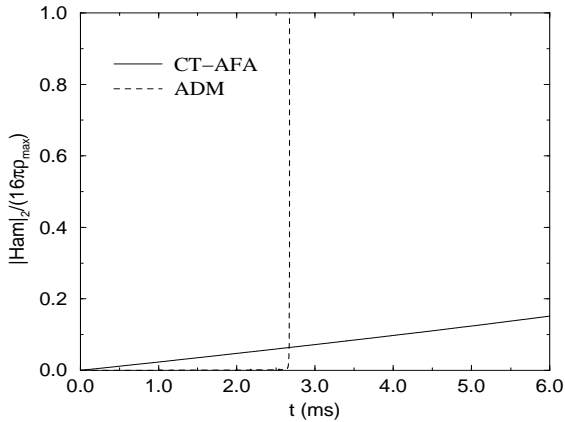} 
\caption{Evolution of the L2-norm of the Hamiltonian constraint for
  a N=1.0 polytropic neutron star model (coupled spacetime and
  hydrodynamical evolution). The ADM system crashes after less than
  2.7ms, while the AFA system evolves stably for a significantly longer
  time. A $64^3$-grid was used to cover the first octant.}
\label{fig:ns_ham1}
\end{figure}

In Fig.~\ref{fig:ns_ham1}, we show the evolution of the L2-norm of the
Hamiltonian constraint for a polytropic, $N=1$, TOV star of
gravitational mass $1.4 M_\odot$ and compactness ratio $M/R=0.146$.  A
$64^3$ grid is used to cover the first octant, with $dx = dy =dz
=0.34$km.  The dashed line corresponds to the ADM system and the solid
line to the AFA system.  Again, as in the vacuum studies, we see that
the ADM evolution suddenly becomes unstable at roughly 2.7ms, while
the AFA evolution remains stable after more than 6ms (we followed
the evolution for more than twice that).

\begin{figure}
\epsfxsize=3.3 in
\epsfysize=2.5 in
\epsfbox{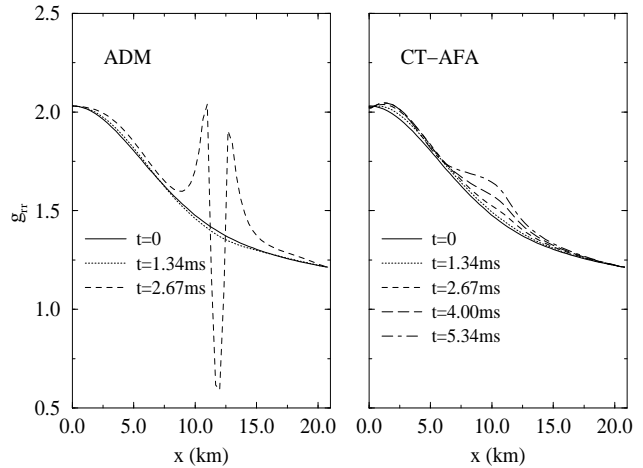}
\caption{
Comparison of the evolution of the radial metric component 
for a $N=1.0$ polytrope with ADM (left panel) and AFA. The evolution
with the latter system proceeds well beyond the time at which the
ADM system becomes unstable.}
\label{fig:ns_grr}
\end{figure}

In Fig.~\ref{fig:ns_grr} we show the evolution of the radial
component of the metric (constructed from the evolved Cartesian metric
components). The first panel of Fig.~\ref{fig:ns_grr} corresponds to
the evolution obtained with ADM.  We see that the star basically
maintains its initial equilibrium, until the high-frequency
instability crashes the code.  In the second panel, we show $g_{rr}$
at various times, obtained with the AFA implementation. All other
parameters are the same as in the ADM evolution.  The ADM run is more
accurate, before it becomes unstable, while the AFA run is stable
but less accurate (there is a secular drift away from the initial
configuration).

The truncation errors of the coupled evolution code initiate a
pulsation of the star in, mainly, its radial modes of pulsation. These
pulsations are damped in time due to the viscosity of the numerical
scheme (see \cite{Stergioulas99,Font99}).  The TVD schemes we are using
describe well the physical pulsations of the fluid, except in a small
region around the center of the star, where short wavelength noise
appears in the radial velocity.  Our trials with other HRSC schemes
show that this behavior seems to be generic for higher order HRSC
schemes~\footnote{We have extensively experimented with other
hydrodynamical evolution schemes. If one uses a first-order (Godunov)
scheme, using piecewise constant reconstructed data for the Riemann
problem, instead of piecewise linear, the radial velocity oscillates
around zero near the center of the star, without any short wave length
noise.  But, a low-order scheme is not capable of accurately
describing the evolution of the stellar surface where the density
distribution is changing rapidly (unless prohibitively large grids are
used) and large errors from the surface layers soon propagate inside
the star. We have also experimented with a mixed system: first-order
near the center and second-order near the surface. In this case the
error grows at the interface of the two regime, yielding a even less
accurate evolution overall.}.  In all such schemes, the radial
momentum near the center has a small residual value of constant sign.
This momentum appears in the r.h.s.  of the evolution equation for
$\tilde \Gamma^i$ (Eq.~(\ref{eq:evolGamma2})).  This, in turn, leads
to an error in the spacetime evolution.  It is noteworthy that this
does not cause an instability in the coupled evolution, except at
very late times, when the violation of the Hamiltonian constraint has
already become extremely large.

\begin{figure}
\epsfxsize=3.3 in
\epsfysize=2.5 in
\epsfbox{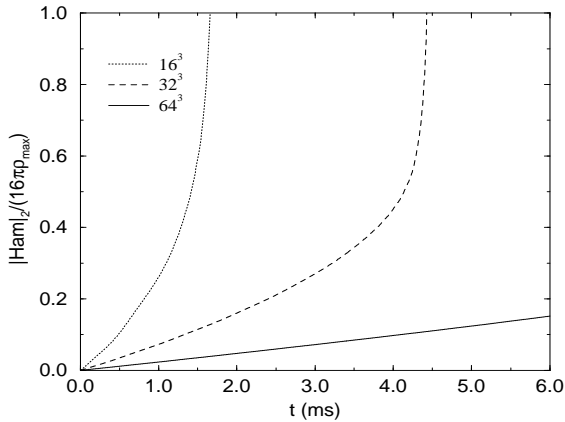}
\caption{Convergence of the L2-norm of the Hamiltonian constraint,
at three different resolutions, for a N=1.0 polytropic
neutron star model. The AFA system is used.}
\label{fig:ns_ham2}
\end{figure}

We note that as the TVD schemes are only first-order accurate at local
extrema, such as the maximum of the density at the center of the star,
so the increase in the Hamiltonian constraint at the center converges
to roughly first order with increasing resolution.  Away from the
center, the scheme is second order convergent.  The convergence of the
L2-norm of the Hamiltonian constraint with the AFA system, for
different grid-sizes (and for the same initial configuration as above),
is shown in Fig.~\ref{fig:ns_ham2}.

\section{Discussion and Conclusions}
\label{sec:conclusions}

In this paper we have studied the stability of three-dimensional
numerical evolutions of the Einstein equations in a formulation that
separates out the conformal and traceless parts of the system. In our
study we have considered different spacetimes including gravitational
waves, black holes, boson stars and neutron stars.

We investigated several implementations of the conformal-traceless
(CT) evolution equations.  We identified two of them which give the
best long term stability behavior: the AF2 implementation for maximal
slicing, and the AFA for algebraic slicings.  The AFA implementation
actively enforces the trace of the conformally rescaled extrinsic
curvature ($\tilde{A}$) to zero at each step of the time evolution,
while the AF2 implementation enforces as well the fact that the trace
of the extrinsic curvature ($K$) should vanish in maximal slicing.  On
the analytic level, the CT evolution equations imply that
$\tilde{A}=0$ throughout the evolution, but this is inevitably
violated in numerically evolution due to truncation error, unless
actively enforced.  Similarly, for maximal slicing, $K$ will not
remain zero numerically unless actively forced to do so.  We find that
these two implementations of the CT equations lead to a more stable
evolution compared to what one can obtain using the standard ADM
evolution equations, under the same resolution, boundary condition and
grid parameter choices, for all systems investigated. In comparison, a
straightforward implementation of the CT equations (``Mom'') is
capable of giving a stable evolution for weak but not strong field
systems.  We should also mention that we have recently become aware of
the work of Lehner, Huq and Garrison~\cite{Lehner00a} where a
comparison of the ADM and CT formulations has been carried out and
where it is also found that freezing the evolution of $K$ (what these
authors call ``locked evolution'') improves considerably the stability
of simulations that use the CT formulation.

Beyond stability, we have also compared the accuracy of the evolutions
obtained by the ADM equations and CT equations.  For all spacetimes
considered we have found that the ADM system is consistently more
accurate than the CT system in short term evolutions, before the
instabilities set in.  Although at present we can offer no explanation
of this difference in accuracy between the different formulations, we
believe that it is not a consequence of our numerical implementation,
but is rather a property of the system of differential equations.
It therefore points in the direction for a possible improvement of the
CT approach.  We note that formulations combining the CT approach and
the hyperbolic approach have been proposed
\cite{Alcubierre99c,Frittelli99}.  A similar investigation of the
stability and accuracy properties of such formulations will be
presented elsewhere.

In this paper we have focused on the implementations and the numerical
properties of their evolutions.  Some understanding of the different
stability of properties on the analytic level is discussed in a
companion paper~\cite{Alcubierre99e}.


\acknowledgements We would like to thank many colleagues for
discussions that have aided the development of this work.  We are
especially grateful to Mark~Miller, Malcolm~Tobias and Wai-Mo~Suen of
the Washington University gravity group, to Toni~Arbona, Carles~Bona,
and Joan~Mass{\'o} of the Universitat de les Illes Balears, to
Gabrielle~Allen, Gerd~Lanfermann and Daniel~Holz at the AEI, and to
Vince~Moncrief.  The research was supported by AEI, NCSA, the NSF
grant Phy 96-00507, NSF NRAC MCS93S025, and NASA HPCC Grand Challenge
Grant NCCS5-153.  J.A.F. acknowledges financial support from a TMR
grant from the European Union (contract nr. ERBFMBICT971902).


\appendix

\section{Stability Analysis of the Iterative Crank-Nicholson Scheme}
\label{sec:ICN}

The numerical scheme used for the simulations described in this paper
is the so-called iterative Crank-Nicholson (ICN) scheme, which is an
iterative, explicit version of the standard implicit Crank-Nicholson
(CN) scheme~\cite{Gustafsson95,New98}. The idea behind this method is
to solve the implicit equations by an iterative procedure, where each
iteration is an explicit operation depending only on previously
computed data.  Normally, this process is stopped after a certain
number of iterations, or until some tolerance is achieved.  For a
linear equation (and in particular in one dimension), the iterative
procedure can easily be much more computationally expensive than the
matrix inversion required to solve the original implicit scheme.  For
a non-linear system, however, solving the implicit scheme directly can
prove to be extremely difficult.

In this appendix we study the stability properties of the ICN scheme
in the particular case of the simple wave equation, and derive two very
important results:

\begin{itemize}

\item In order to obtain a stable scheme one must do {\em at least} three
iterations, and not just the two one would normally expect (two
iterations are enough to achieve second order accuracy, but they are
unstable!).

\item The iterative scheme itself is only convergent if the standard
Courant-Friedrichs-Lewy (CFL) stability condition is satisfied,
otherwise the iterations diverge.

\end{itemize}

These two results taken together imply that there is no reason (at
least from the point of view of stability) to ever do more that three
ICN iterations.  Three iterations are already second order accurate,
and provide us with a (conditionally) stable scheme.  Increasing the
number of iterations will not improve the stability properties of the
scheme any further.  In particular, we will never achieve the
unconditional stability properties of the full implicit CN scheme,
since if we violate the CFL condition the iterations will diverge.
\footnote{As we were finishing this manuscript we became aware of a
  paper by S.~Teukolsky were he does essentially the same analysis and
  obtains the same results~\cite{Teukolsky99}.  His analysis and ours
  complement each other, since he considers any finite number of
  iterations, while we consider only 1, 2 and 3 iterations. On the
  other hand, here we also consider the question of the convergence
  properties of an {\em infinite} number of iterations.}

For our stability analysis we will consider the simple wave equation
in N-dimensions.  Numerical experiments have shown that the full
Einstein equations have essentially the same stability properties.

Consider then the N-dimensional wave equation written in ``3+1 like'' form:
\begin{equation}
\partial_t \phi = A , \qquad \partial_t A = \sum_{i=1}^N \partial_i^2 \phi .
\end{equation}

For the finite difference approximation to these equations we employ
the usual notation
\begin{equation}
f^n_{\bf m} := f(x_i=m_i \Delta x,t=n \Delta t) ,
\end{equation}

\noindent with $n$ and ${\bf m}=(m_1,...,m_N)$ integers.  The implicit CN
scheme is then given by
\begin{eqnarray}
\phi^{n+1}_{\bf m} &=& \phi^n_{\bf m} + \frac{\Delta t}{2}
\left( A^{n+1}_{\bf m} +  A^{n}_{\bf m} \right) , \\
A^{n+1}_{\bf m} &=& A^{n}_{\bf m} + \frac{\Delta t}{2 (\Delta x)^2}
\sum_{i=1}^N \delta^2_i \left( \phi^{n+1}_{\bf m} +  \phi^{n}_{\bf m}\right) ,
\end{eqnarray}

\noindent where the finite difference operators $\delta^2_i$ are defined as
\begin{equation}
\delta^2_i f^n_{m_i} := f^n_{m_i+1} - 2 f^n_{m_i} + f^n_{m_i-1} .
\end{equation}

The implicit CN scheme is well known to be unconditionally stable for
the wave equation (i.e. stable for any value of $\Delta t$).

The ICN scheme is defined in the following way
\begin{eqnarray}
\phi^{(1)}_{\bf m} &=& \phi^n_{\bf m} + \Delta t \;  A^{n}_{\bf m} , \\
A^{(1)}_{\bf m} &=& A^{n}_{\bf m} + \Delta t
\sum_{i=1}^N \phi^n_{\bf m} , \\
\nonumber \\
\phi^{(i)}_{\bf m} &=& \phi^n_{\bf m} + \frac{\Delta t}{2}
\left( A^{(i-1)}_{\bf m} +  A^{n}_{\bf m} \right) , \\
A^{(i)}_{\bf m} &=& A^{n}_{\bf m} + \frac{\Delta t}{2 (\Delta x)^2}
\sum_{i=1}^N \delta^2_i \left( \phi^{(i-1)}_{\bf m} + \phi^{n}_{\bf m}\right) ,
\end{eqnarray}
and finally,
\begin{eqnarray}
\phi^{n+1}_{\bf m} &=&  \phi^{(i_{\rm max})}_{\bf m} , \\
A^{n+1}_{\bf m} &=&  A^{(i_{\rm max})}_{\bf m} ,
\end{eqnarray}

From these expressions it is clear that if the iterations converge, we
will recover the implicit CN scheme.

For the stability analysis of the ICN scheme we use the standard von
Neumann ansatz~\cite{Gustafsson95,Press86}
\begin{eqnarray}
\phi^n_{\bf m} &=& \xi_1 \lambda^n e^{i ({\bf k} \cdot {\bf m}) \Delta x} , \\
A^n_{\bf m} &=& \xi_2 \lambda^n e^{i ({\bf k} \cdot {\bf m}) \Delta x} ,
\label{eq:vonNewmann}
\end{eqnarray}

\noindent with ${\bf k}$ the ``wave vector''.  Notice that the highest wave
number that can be represented on the finite difference grid
corresponds to $k_i \Delta x = \pi$.  The stability condition for our
numerical scheme will then be
\begin{equation}
| \lambda | \leq 1 .
\label{eq:stability}
\end{equation}

Let us consider first the ``1-step'' ICN scheme, that is, the
so-called forward-time centered-space (FTCS) scheme.  This scheme is
well known to be only first order accurate, and unconditionally
unstable.  The fact that is only first order accurate can be easily
seen from a simple Taylor expansion in time.  For the stability
analysis we substitute the von Neumann ansatz~(\ref{eq:vonNewmann}) into
the ICN scheme defined above with \mbox{$i_{\rm max} = 1$}.  Doing
this we obtain
\begin{equation}
\lambda^2 - 2 \lambda + 1 + 2 \rho^2 u^2 = 0 ,
\end{equation}
where $\rho := \Delta t / \Delta x$ is the Courant parameter and
\begin{eqnarray}
u^2 &:=& \sum_{i=1}^N u_i^2 , \\
u_i^2 &:=& 1 - \cos (k_i \Delta x) .
\end{eqnarray}

Solving for $\lambda$ we find
\begin{equation}
\lambda = 1 \pm i \sqrt{2} \; \rho u ,
\end{equation}
which implies
\begin{equation}
| \lambda | = 1 + 2 \rho^2 u^2 > 1 .
\end{equation}
Comparing with~(\ref{eq:stability}) we conclude that the 1-step scheme
is unstable for any value of $\Delta t$.

Let us now consider the 2-step scheme.  If we take the ICN scheme
above with $i_{\rm max}=2$, and do the appropriate substitutions we
find
\begin{eqnarray}
\phi^{n+1}_{\bf m} &=& \phi^{n}_{\bf m} + \Delta t \; A^{n}_{\bf m}
+ \frac{\rho^2}{2} \sum_{i=1}^N \delta^2_i \; \phi^{n}_{\bf m} , \\
A^{n+1}_{\bf m} &=& A^{n}_{\bf m} + \frac{\rho^2}{2}
\sum_{i=1}^N \delta^2_i \; \left( 2 \phi^{n}_{\bf m}
+ \Delta t A^{n}_{\bf m} \right) .
\end{eqnarray}
As before, a simple Taylor expansion shows that this approximation is
now second order both in time and space.

Using again the ansatz~(\ref{eq:vonNewmann}) we find now that
\begin{equation}
\lambda^2 + 2 \lambda \left( \rho^2 u^2 -1 \right) + 1 + \rho^4 u^4 = 0 .
\end{equation}

Solving again for $\lambda$ we obtain
\begin{equation}
\lambda = 1 - \rho^2 u^2 \pm i \sqrt{2} \; \rho u ,
\end{equation}
which implies
\begin{equation}
| \lambda | = 1 + \rho^4 u^4 > 1 .
\end{equation}

Comparing again with~(\ref{eq:stability}) we conclude that the 2-step
ICN scheme is also unstable for any value of $\Delta t$.  This result
is surprising, since a priori one might expect that the 2-step scheme
should behave like a predictor-corrector scheme, and should therefore
be stable.

Finally, let us consider the 3-step scheme. By taking the ICN scheme
above with $i_{\rm max}=3$, and doing the appropriate substitutions we
now find
\begin{eqnarray}
\phi^{n+1}_{\bf m} &=& \phi^{n}_{\bf m} + \Delta t A^{n}_{\bf m}
+ \frac{\rho^2}{4} \sum_{i=1}^N \delta^2_i \left( 2 \phi^{n}_{\bf m}
+ \Delta t A^{n}_{\bf m} \right) , \\
A^{n+1}_{\bf m} &=& A^{n}_{\bf m} + \frac{\rho^2}{2}
\sum_{i=1}^N \delta^2_i \; \left( 2 \phi^{n}_{\bf m}
+ \Delta t A^{n}_{\bf m} \right) \nonumber \\
&& + \frac{\rho^3}{4 \Delta x}
\left( \sum_{i=1}^N \delta^2_i \right)^2 \phi^{n}_{\bf m} .
\end{eqnarray}
A Taylor expansion now shows that this 3-step scheme is still only
second order accurate in both time and space.  Using the
ansatz~(\ref{eq:vonNewmann}) on this scheme we now find
\begin{equation}
\lambda^2 + 2 \lambda \left( \rho^2 u^2 -1 \right) + 1 - \rho^4 u^4 
+ \frac{1}{2} \rho^6 u^6 = 0 .
\end{equation}

And solving for $\lambda$ we obtain
\begin{equation}
\lambda = 1 - \rho^2 u^2 \pm i \sqrt{2} \; \rho u \left| 1 - \rho^2 u^2 /2
\right| ,
\end{equation}
which now implies
\begin{equation}
| \lambda | = 1 - \rho^4 u^4 + \frac{1}{2} \rho^6 u^6 .
\end{equation}

Comparing now with~(\ref{eq:stability}) we obtain the following
stability condition
\begin{equation}
\rho^2 u^2 \leq 2 .
\end{equation}
\noindent And finally, from the fact that the maximum value of $u^2$ is
$2\sqrt{N}$ we find
\begin{equation}
\rho \leq 1 / \sqrt{N} .
\end{equation}
Notice that this is just the standard CFL condition in $N$ dimensions.
We then conclude that in order to obtain a (conditionally) stable
scheme we need to do at least three iterations.

Next, we address the question of the stability of the iterations
themselves, that is, if we iterate an infinite number of times do we
converge to something (that is, to the implicit CN scheme)?  For this
we consider two consecutive iteration steps $(i-1,i)$, and subtract
them to get
\begin{eqnarray}
\phi^{(i)}_{\bf m} - \phi^{(i-1)}_{\bf m} &=& \frac{\Delta t}{2} \left(
A^{(i-1)}_{\bf m} - A^{(i-2)}_{\bf m} \right) , \\
A^{(i)}_{\bf m} - A^{(i-1)}_{\bf m} &=& \frac{\Delta t}{2 (\Delta x)^2}
\sum_{i=1}^N \left( \phi^{(i-1)}_{\bf m} - \phi^{(i-2)}_{\bf m}\right) .
\end{eqnarray}

Let us now define \mbox{${F_1}^{(i)}_{\bf m} := \phi^{(i)}_{\bf m} -
\phi^{(i-1)}_{\bf m}$} and \mbox{${F_2}^{(i)}_{\bf m} := A^{(i)}_{\bf m} -
A^{(i-1)}_{\bf m}$}.  The above equations become
\begin{eqnarray}
{F_1}^{(i)}_{\bf m} &=& \frac{\Delta t}{2} {F_2}^{(i-1)}_{\bf m} , \\
{F_2}^{(i)}_{\bf m} &=& \frac{\Delta t}{2 (\Delta x)^2} \sum_{i=1}^N
{F_1}^{(i-1)}_{\bf m} .
\end{eqnarray}

We now use the von Neumann ansatz again
\begin{eqnarray}
{F_1}^{(i)}_{\bf m} &=& f_1 \lambda^i e^{i ({\bf k} \cdot {\bf m})
\Delta x} , \\
{F_2}^{(i)}_{\bf m} &=& f_2 \lambda^i e^{i ({\bf k} \cdot {\bf m})
\Delta x} ,
\end{eqnarray}

Substituting this ansatz back into the equations above we find
\begin{equation}
\lambda^2 + \frac{1}{2} \rho^2 u^2 = 0 ,
\end{equation}
from which we obtain
\begin{equation}
\lambda = \pm i \frac{\rho u}{\sqrt{2}} .
\end{equation}

In this case, the condition for the iterations to converge implies
that the norm of the successive differences should go to zero, which in
turn implies \mbox{$| \lambda | < 1$}.  Using again the fact that the
maximum value of $u^2$ is $2 \sqrt{N}$ we see that the convergence
condition reduces to
\begin{equation}
\rho < 1 / \sqrt{N} .
\end{equation}

This is again the standard CFL stability condition.  So we have just
shown that if this condition is violated, the iterations will fail to
converge.  This means that there is no reason to try to iterate to
convergence in the hope of improving stability.  If $\Delta t$ was too
big in the first place the iterations will never converge.


\bibliographystyle{prsty}

\bibliography{bibtex/references}

\end{document}